\documentclass[11pt]{article}               
\usepackage{a4}
\usepackage{epsfig}
\usepackage{amsmath,amssymb}

\begin{document}

\begin{flushright} 
\vspace*{-30mm}
{\bf LAL 03-59}\\
November 2003\\
\end{flushright}
 
\vspace{15mm}
\begin{center}
{\Large \bf PHYSICS OF THE LINEAR COLLIDER}\\

\vspace{10mm}
{\large \bf Fran\c cois Richard}\\

\vspace{5mm}
{\bf Laboratoire de l'Acc\'el\'erateur Lin\'eaire,} \\
IN2P3-CNRS et Universit\'e de Paris-Sud, B\^at. 200, BP 34 - 91898 Orsay Cedex, France \\
\vspace{10mm}
{\it  Invited talk given at the XXIst International Symposium on Lepton and Photon Interactions \\
\it at High Energies, 11-16 August 2003, FNAL, Batavia, Illinois, USA }
\end{center}


\vspace{5mm}
\begin{abstract}
This presentation intends to illustrate the specific capabilities
 of an e$^+$e$^-$ sub-TeV collider to provide answers on the basic issues in physics: origin of mass, hierarchy of masses, cosmological problems. Some foreseeable scenarios are discussed with a possible synergy with LHC.
\end{abstract}

\vspace{10mm}

\baselineskip=13.07pt
\section{Introduction}
In this presentation I intend to summarize the main physics prospects
of the Linear
Collider (LC) presently under consideration in North America, Asia and Europe. These
prospects have been studied by 3 communities and there exist various documents describing them in detail\cite{web}.\par
After recalling briefly the baseline for the future LC, I will mention some important features of the detectors under consideration and the requirements 
needed for the machine parameters. \par
Concerning the physics, I will focus on 3 aspects:     

\begin{enumerate}
\item
The mechanism of electro-weak symmetry breaking (EWSB), in other words what is the origin of mass in particle physics. This aspect will be my main emphasis. 
\item
The problem of mass hierarchy, in particular in the Higgs sector, and the need
for new mechanisms beyond the Standard Model (SM) like Supersymmetry (SUSY).  
\item
The input on cosmology of this model which can explain the origin of Dark Matter in the universe.
\end{enumerate}
In a short presentation one can only give a very partial view of ongoing studies performed in the 3 
regions. Examples of uncovered or very partially covered topics are: e$^+$e$^-$ physics at MultiTeV (CLIC scheme\cite{clic}), 
the various SUSY breaking scenarios, SUSY and CP violation,
SUSY and the neutrino sector, extra dimensions with different schemes either alternate or combined with SUSY, e$^-$e$^-$,$\gamma$e and $\gamma\gamma$ physics, precise test of QCD with a LC.

\newpage
  
\section{The TeV LC}
The present goal is to construct an e$^+$e$^-$ LC covering an energy between the Z boson mass and 500 GeV, with 
polarized electrons (at least 80\%) and collecting 500 fb$^{-1}$ in the first four years of running. 
This LC should
be able to reach, in a second stage, an energy $\sim$ 1 TeV, collecting about 500
 fb$^{-1}$/year. \par
Various options are considered, which would require additional equipments and whose priorities will depend on the physics priorities emerging after LHC and LC operation:
 
\begin{enumerate}
\item
Positron polarization, at the 60\% level, not easy to implement but needed to fully exploit the precision of
a Z factory (the GigaZ scheme). This polarization is also needed for transverse polarization measurements which have recently been emphasized\cite{Rizzo}.  
\item
An  e$^-$e$^-$    collider, possible with reduced hardware changes 
\item
A  $\gamma\gamma$ (or a $\gamma$e collider), which could operate with a maximum energy $\sim$ 80\% of the nominal energy. This scheme  
would require major changes in the interaction region. 
As for the e$^-$e$^-$ case, there will be a reduction in luminosity at the same energy but a high degree of polarization is feasible. Contrary to  e$^+$e$^-$, a zero crossing angle is not possible even with the supraconductive technology.
\end{enumerate}   

\section{The Detector}
Given the LC luminosity, two orders of magnitude above LEP, in several analyses the precision will be 
limited by systematic uncertainties. Part of these uncertainties come from detector limitations but this will 
improve using better technologies:
   \begin{enumerate}
\item
Improved vertexing which will allow a clean and efficient separation of charm quarks and also tagging of
tau leptons. The improvement with respect to LEP is due to a small radius beam pipe and to the resolving power of 
thin pixel detectors.    
\item
Improved energy flow, the aim is to improve the jet energy resolution by about a factor 2 with respect to LEP/SLD with fine segmentation of the calorimeters which are inside the magnetic coil. This improvement should allow to cope efficiently with 6/8 jet topologies from ZHH and ttH channels. 
This resolution will also allow a clean separation between ZZ and WW hadronic final states to isolate the
WW$\nu\bar{\nu}$ channel.
\item Momentum resolution will be improved by a factor 10 with respect to LEP/SLD, with polar angle coverage down to 
100 mrad.
\item Hermeticity on energetic $\gamma$ and electrons should go down to about 5 mrad with instrumented masks
and good segmentation of the very forward calorimeter. 
\end{enumerate}
A detailed discussion of the new technologies implied can be found in the presentation given by T. Behnke at this conference. \par
A limiting factor will also come from our  knowledge of the differential luminosity, of the polarization and
of the energy calibration. Physics groups and machine experts are actively investigating these issues.

\section{Which physics scenario for EWSB ?}
This is clearly the central issue for LHC and LC. While there is no doubt that LHC should discover a SM/MSSM type Higgs boson, one should be prepared to unconventional answers from Nature and I will illustrate this schematically in the following sections.  \par
From LEP/SLD/TeVatron precision measurements (PM) one can derive 2 important consequences:
   \begin{enumerate}
\item The Standard Model (SM) or its supersymmetric (SUSY) minimal extension (MSSM) are compatible with PM.
\item Unification of the 3 interactions occurs at $\sim$2 10$^{16}$ GeV provided that the MSSM mass spectrum is at
$\sim$1 TeV (see figure 1 taken from\cite{zerw}). 
\end{enumerate}    

\begin{figure}[h]
\begin{center}
\psfig{figure=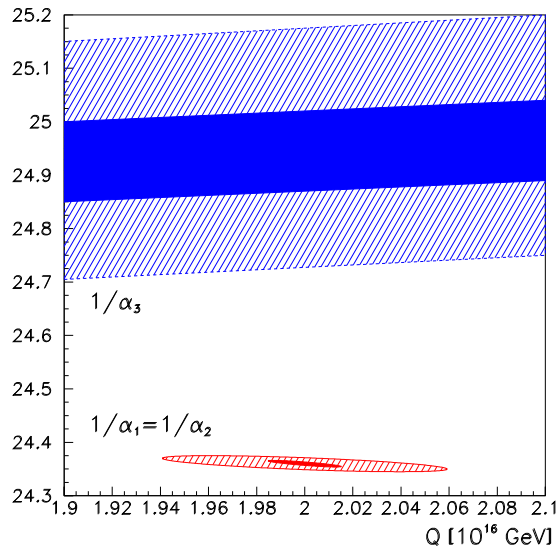,width=8truecm}
\caption{This figure shows that while LEP/SLD data  (hatched) are compatible with 
SUSY-GUT, there is an indication of a discrepancy which, if real, should become clear with the
precision of GigaZ(full).}
\end{center}
\end{figure}

A light Higgs is therefore expected, below 130 GeV as a consequence of MSSM, below 250 GeV as a consequence of PM assuming that there is no contribution other than SM. \par 
With a critical view\cite{chano} of PM, one can notice that sin$^2\theta_W$ from the charge asymmetry on b quarks measured at LEP1 
is hardly consistent with the value obtained at SLD from polarization asymmetry\cite{PM}. So far no compelling explanation has emerged from theory nor experiment. New physics interpretations, although not impossible\cite{wagner}, are severely 
limited by the absence of a significant deviation on the b quark cross-section. \par
Another discrepancy, found in NuTeV, is still under investigation, but it should be noticed that this determination of sin$^2\theta_W$ has no sensitivity to the value of the Higgs boson mass\cite{PM}. \par
When removing from the fit the b quark asymmetry result one finds that the Higgs mass value indicated by
the data is too low given the direct search limit. At this stage one can remark that the effect is at the 2 sigma level and that the recent update on the top mass found by D0 with run 1 data and the shift in the W mass after a reanalysis of ALEPH data should even reduce it further.\linebreak 
\noindent
In our opinion one should therefore wait for the top mass new measurements at FNAL before drawing any definite conclusion on this discrepancy. \par
Another possibility could be that there are non SM/MSSM contributions coming from an alternate 
scheme (see\cite{Peskinw} for a general discussion). I will therefore consider, for simplicity, that a LC  could be dealing with three  EWSB scenarios:
\begin{enumerate}

\item There is a light Higgs boson consistent with MSSM. There will then be emphasis on the precise measurement of Higgs couplings based on the collection of $\sim$10$^5$ HZ events.
\item  There is a Higgs boson but with a mass incompatible with SM/MSSM. A LC would then focus its effort 
in  understanding the underlying mechanism and search for direct or indirect signals of new physics.
\item  There is no Higgs boson, implying that the longitudinally polarized gauge bosons will strongly 
interact.
A LC would then run at its maximal energy and focus on WW final states.
\end{enumerate}  
I will illustrate with some examples how a sub-TeV LC can provide the appropriate answers in these three
scenarios.\\

\vspace{5mm}
\begin{figure}[h]
\begin{center}
\psfig{figure=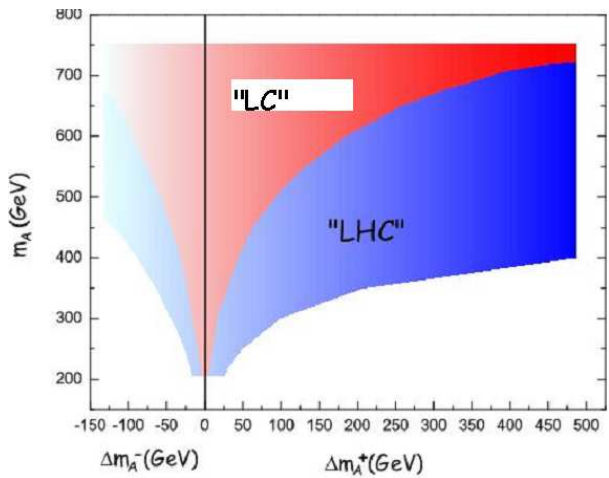,width=8.truecm}
\vspace{5mm}
\caption{Sensitivity to the pseudoscalar Higgs boson mass at LC and LHC. This figure
assumes an integrated luminosity of 300 fb$^{-1}$ and the combination of 2 experiments for LHC.} 
\end{center}
\end{figure}

\subsection{MSSM scenario}
The first questions are obviously: are we dealing with a CP even scalar, are the coupling to fermions and to bosons consistent with MSSM ? 
   \begin{enumerate}
\item A LC will identify unambiguously the spin-parity of a Higgs boson by measuring the shape of the HZ cross-section near threshold an the angular distributions provided by the ZH channel.
\item All fermionic (except for the top) and bosonic couplings will be measured at the \% level for tree level couplings, 
at 5\% for the gluonic width $\Gamma_{gg}$ and 20\% for  $\Gamma_{\gamma\gamma}$. For the later one can reach the \% level using the photon collider scheme.
\item The Higgs coupling to the top quark can be measured with an 7-15\% accuracy in the \linebreak 120-200 GeV mass range. For the Higgs self-coupling $\lambda_{HHH}$ the accuracy would be 10\% (20\%) at 800 GeV (500 GeV)center of mass energy.        
\end{enumerate}
With such accuracies one can detect with high sensitivity the presence of a non-standard component:

 \begin{enumerate}
\item In the MSSM itself, it is possible to detect the influence of the heavier Higgs bosons. As shown in figure 2, one can estimate\cite{gross} the mass of these bosons up to about 700 GeV. This goes beyond the mass reach of an  e$^+$e$^-$ LC but,
given this information, could motivate a photon collider scheme in which these Higgs bosons can be singly produced.

\begin{figure}[h]
\begin{center}
\vspace{-8mm}
\psfig{figure=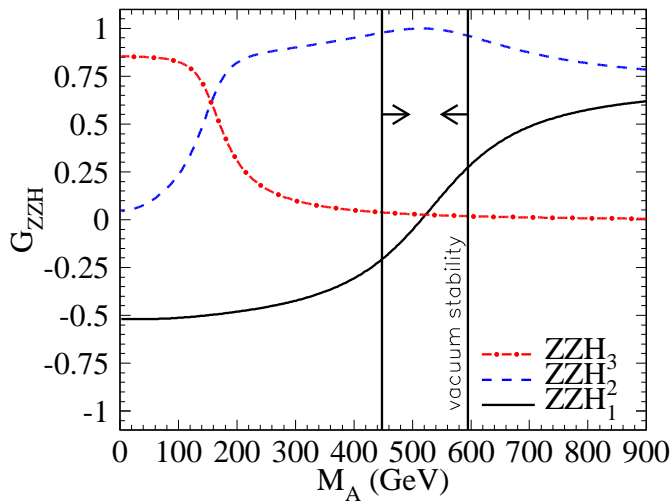,width=10.5truecm}
\caption{Variation of the ZZH coupling with the mass of the pseudoscalar A, in an NMSSM model, for the 3 scalar states.}
\end{center}
\end{figure}

\item Beyond MSSM one could have CP violation in the Higgs sector or NMSSM with an additional Higgs isosinglet which can mix with the isodoublets. This could result in a significant drop in the ZZH coupling as 
shown\cite{miller} in figure 3. In this respect a LC is very robust and can stand a reduction of a factor 100 in the ZH cross-section.

\item Similarly it is worth recalling that the Higgs detection does not depend on the final state branching ratios, in particular LC can very well detect an 'invisible' Higgs, which may occur in various schemes\cite{invis}. Moreover if $\Gamma_{inv}$ is at the 2\% level, the LC could still give a 5 standard deviation evidence on the presence of an invisible channel. An example\cite{boud} is given within SUSY where one assumes an unusual hierarchy between gaugino masses M1 and M2 
(M1$\sim$M2/10), such that the limit from LEP2 on the chargino does not eliminate the possibility of a very light LSP. Figure 4 shows the large effects possible in this scheme.
\newpage

\begin{figure}[h]
\begin{center}
\psfig{figure=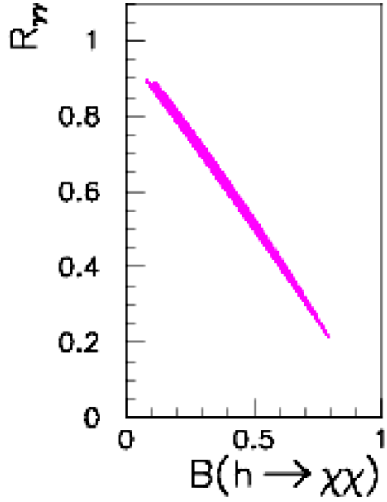,width=5truecm}
\vskip 3mm
\caption{Variation of R$_{\gamma\gamma}$, the rate at LHC of Higgs into a pair of photons normalized to the SM rate,
versus the branching ratio of the Higgs into a pair of LSP in a model with non universal gaugino masses.}
\end{center}
\end{figure}

\item Theories with large extra dimensions also provide valid schemes for EWSB. Within these theories the radion is a scalar field introduced to stabilize the small dimensions. It can therefore mix with the Higgs boson and, accordingly, modify the couplings of the Higgs to bosons and fermions as seen in figure 5. This 
figure\cite{hewett} shows that the modification is similar for W and fermions and therefore independent on the ratio. A LC however has access to the couplings and would measure this variation. The gluon pair BR should also provide a valid input in the interpretation of this effect.
\end{enumerate}

\vspace{5mm}

\begin{figure}[h]
\begin{center}
\vspace{-5mm}
\psfig{figure=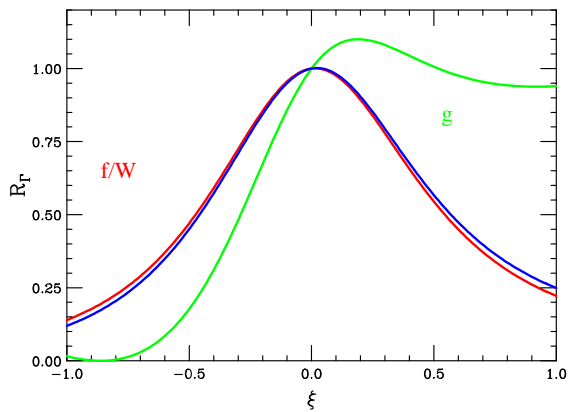,width=8.truecm}
\end{center}
\caption{Ratio of the Higgs widths to their SM value for  fermions, W bosons and gluons  as a function  of the  mixing  parameter in  a radion model, assuming a Higgs  mass of  125 GeV. }

\end{figure}

\vspace{-5mm}

\subsection{Quantum level consistency}

The accuracy on the indirect determination of the Higgs mass can be improved by an order of magnitude with respect to LEP1/SLD. GigaZ should measure sin$^2\theta_W$ with an error $\sim$10$^{-5}$ provided one has polarized positrons. Figure 6 shows the other limiting accuracies and how they should evolve.

\begin{figure}[h]
\begin{center}
\psfig{figure=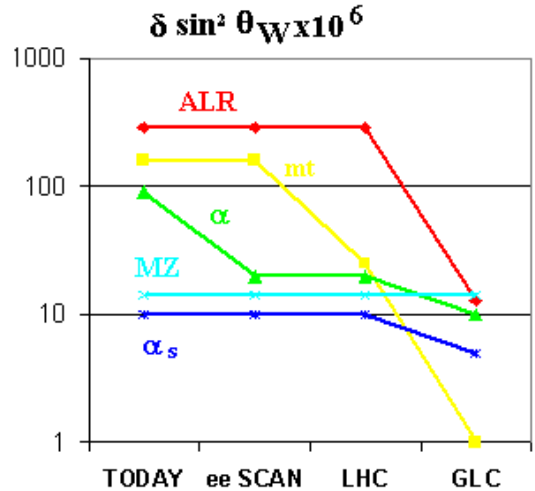,width=7.5truecm}
\caption{Predicted evolution of the experimental and theoretical accuracy on sin$^2\theta_W$ with the expected progress
on the polarization asymmetry, the top mass measurement, the decrease in uncertainty on $\alpha$(M$_Z$)
provided by e$^+$e$^-$ scans and the improved precision on $\alpha_s$ at GigaZ.}
\end{center}
\end{figure} 
\vspace{-3mm}
\noindent
 The B and K factories should provide the
input needed to settle the issue on $\alpha(M_Z)$ while the top mass accuracy with a LC should eliminate this
source of error. \par

In parallel, there is a continuous and fruitful effort in reducing the theoretical uncertainty which is now coordinated in the 'Loopverein' working group\cite{loop}. As an example, figure 7 indicates the continuous 
progress\cite{jadach} achieved on theoretical errors for the luminosity.

\begin{figure}[h]
\begin{center}
\vspace{-2mm}
\psfig{figure=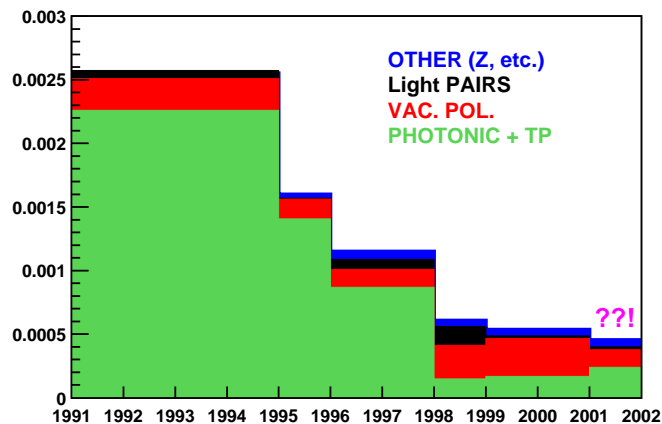,width=9.5truecm}
\caption{Components of the luminosity theoretical error at LEP1.}
\end{center}
\end{figure} 

\noindent
For what concerns sin$^2\theta_W$, the present theoretical uncertainty\cite{hollik}, $\sim$ 610$^{-5}$, has to be reduced to match the expectations given in figure 6.\par
With GigaZ one could therefore test at the 5\% level the equality:

$$ M_HDirect=M_HIndirect $$

and draw valid conclusions on the global consistency of SM or MSSM. \par

An alternate to GigaZ, if positron polarization is not available, would be an improved measurement of the W 
mass using a threshold scan. The critical issue, there, is energy calibration: one has to extrapolate the beam measurement 
from
the Z mass to the W threshold with a precision at $\sim$510$^{-5}$. One expects $\delta$MW$\sim$6 MeV, which gives a 10\% accuracy on the indirect Higgs boson mass.
\subsection{non-MSSM Higgs scenario}
Let us assume that a Higgs boson has been found with a mass inconsistent with PM, meaning above 200 GeV. 
This mass is therefore also inconsistent with MSSM. While PM are still constraining the gauge sector since, for instance, one could measure the BR of the Higgs into WW at 5\% with \linebreak Mh=250 GeV, the main mission of a LC would then be to find the 'guilty part'. \par

If there is direct evidence for new physics at LHC, e.g. if one observes a candidate Z', LC can decipher the
message with precision measurements in the channel  e$^+$e$^-$ $\rightarrow f\bar{f}$. \par
With measurements in the TeV range, LC provides high discrimination against the various Z' predicted by several symmetry groups\cite{tdrt} as shown in figure 8. Knowing the Z' mass from LHC, LC PM at high energy will extract the vector and axial couplings of this Z'. \par


\begin{figure}[h]
\centering
\vspace*{1.mm}
\psfig{figure=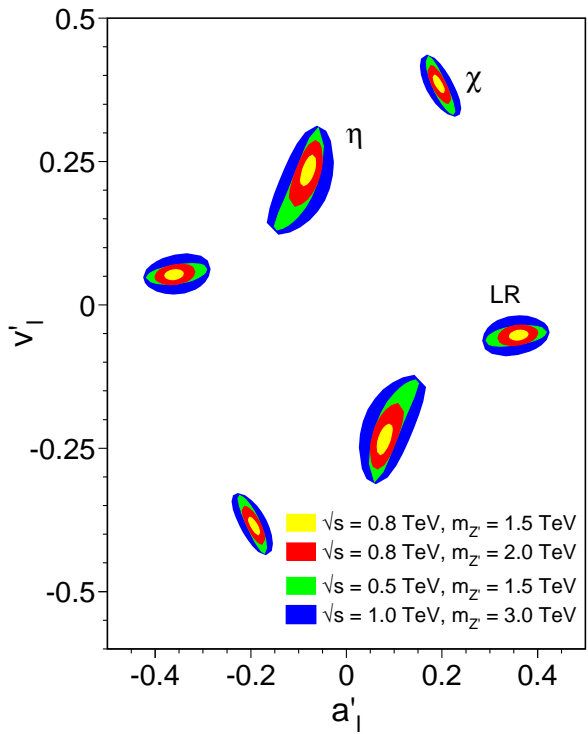,width=8truecm}
\vspace*{-1mm}
\caption{Accuracies expected on Z' vectorial and axial couplings from a LC operating up to 800 GeV for various extensions of the SM.}
\end{figure}

\newpage
Similarly one can use GigaZ to measure Z-Z' mixing, providing extra informations\cite{fr}. These measurements should also allow to restore the consistency of PM with the observed Higgs mass. \par
As shown in figure 9 the mass domain of a LC covers\cite{fr} and, in some instances, surpasses, the LHC domain. It may
therefore turn out that only the LC information will be left to solve the puzzle of a Higgs mass inconsistent with PM.

\begin{figure}[h]
\centering
\vspace{-5mm}
\psfig{figure=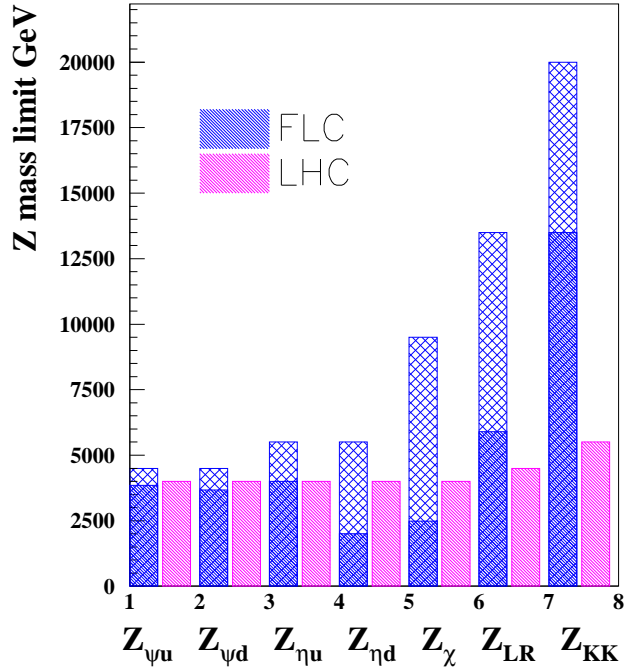,width=10truecm}
\caption{Mass domain covered by LHC (pink) and LC (blue) for various scenarios predicting a Z'. In the case of
LC the dark hatched part corresponds to the limit of sensitivity through mixing, while the light hatched part corresponds to the sensitivity through interference for a LC operating at 800 GeV.}
\end{figure}

\par
In some cases the Z' can decouple from ordinary fermion as in the UED model (Universal Extra Dimensions\cite{appel})  
where the new particles are mass degenerate Kaluza Klein excitations of ordinary particles which carry a conserved quantum number and therefore need to
be pair-produced. Remarkably, given the absence of direct coupling to fermions, this model has much weaker mass bounds than usual extensions with new Z' or extra dimensions and therefore it is not yet excluded that 
there could be pair-production at a TeV LC. If not, 
 there still remains the effect on weak isospin violation in the top sector which contributes to the $\rho$ parameters and therefore is measurable at GigaZ. 

\subsection{The Little Higgs scenario}

Among the possible non-MSSM extensions of SM, the 'Little Higgs' model\cite{lh} has received special attention since it offers a viable solution to the hierarchy problem in the Higgs sector. In this scheme, the Higgs boson is a pseudo-goldstone boson originating from a symmetry broken at 10-100 TeV. There is a perturbative theory below this scale and 
quadratic divergences of the Higgs mass are cancelled at first order by the contribution of new particles originating from the new symmetries. In particular there could be a light
U(1) gauge boson, the B', which contributes
to the $\rho$ parameter in the minimal model called 'Littlest Higgs'. \par
If LHC finds this B' then, as stated previously, LC will allow to identify its origin. If not, LC can predict\cite{fr} 
the mass of this object from PM and indicate which improvements in luminosity/energy are needed at LHC (or at future colliders) to discover it. \par 
To conclude on this non-MSSM Higgs scenario, one can say that it shows in a quantitative form how there could be a strong 
synergy between LC and LHC and that it suggests that we may need the full information of a LC from the Z pole to the maximum energy.

\vspace{5mm}
\subsection{No Higgs scenario}

In the case of no Higgs or, equivalently of a very heavy Higgs boson, with a mass above 1 TeV, one needs, 
even more than for the previous scenario, some kind of conspiracy cancelling the contribution of the Higgs 
term in PM. \par

The most direct manifestation of the heavy Higgs should occur in the gauge sector, where longitudinally polarized W should strongly interact as a result of the absence of a Higgs exchange term cancelling the divergence in the process W$_L$W$_L\rightarrow$ W$_L$ W$_L$. \par
A likely scenario would be that the
strong interaction, as in QCD, results into a $\rho$-type resonance with a mass below $\Lambda_{EWSB}=4\pi v\sim$3 TeV. Using the channel   e$^+$e$^-$ $\rightarrow$W$^+$W$^-$ and extracting the longitudinal part, one can observe a clear 
effect\cite{bark} with a LC operating at 800 GeV as shown in figure 10.

\begin{figure}[h]
\centering
\vspace{3mm}
\psfig{figure=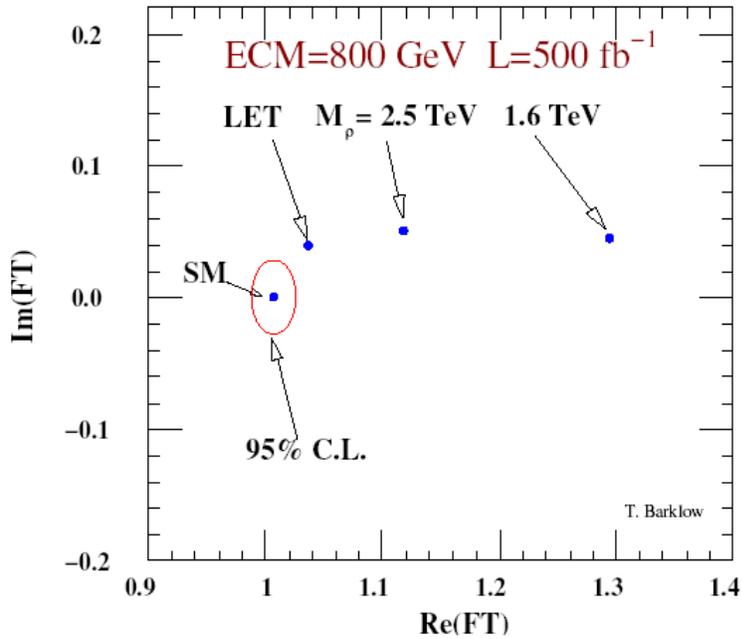,width=12truecm}
\caption{Sensitivity on the W$_L$ form factor FT with a LC operating at 800 GeV and an integrated luminosity of 500 fb$^{-1}$ when there is a $\rho$ resonances at 1.6 TeV or 2.5 TeV or, with no resonance, the effect predicted the Low Energy Theorem.}
\end{figure} 
\newpage
\noindent
Figure 11 displays the energy 
dependence of the signal if there is a resonance. Even if there is no resonance (figure 10),  one should still observe a minimal effect, the so-called LET term predicted on a model-independent basis (Low Energy Theorem).

\begin{figure}[h]
\centering
\psfig{figure=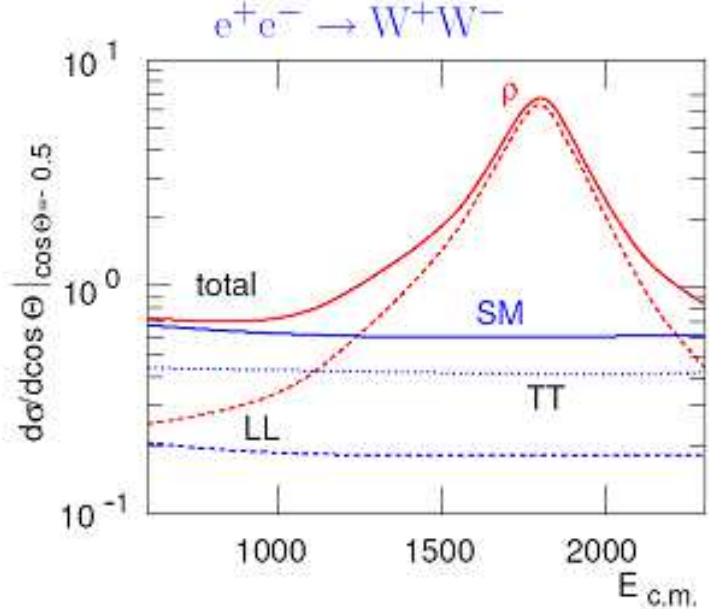,width=11truecm}
\caption{Variation with the LC energy of the backward differential 
cross-section for W$^+$W$^-$ with a $\rho$ resonance. The dashed read curve corresponds to the LL component part to be compared to the dashed curve blue curve corresponding to the SM.}
\end{figure} 

\vspace{3mm}
\noindent
 Table 1 summarizes the 
evidence foreseen at LC and LHC. One can clearly see an advantage for LC if there is a resonance, while the 2 machines are roughly equivalent otherwise. \par

\vspace{3mm}
\begin{table}[h]
\begin{center}

\begin{tabular}{|c|c|c|c|c|}
\hline
&
{$\sqrt{s}$ GeV} &
{$\mathcal{L}$ fb$^{-1}$} &
{M$_\rho$ 1.6 TeV} & LET \\
\hline
LC &
0.5 &
300 &
16 $\sigma$ & 3 $\sigma$ \\
\hline LC & 0.8 & 500 & 38 $\sigma$ & 6 $\sigma$ \\
\hline LC & 1.5 & 200 & 204 $\sigma$ & 5 $\sigma$\\ 
\hline LHC & 14 & 100 & 6 $\sigma$ & 5 $\sigma$ \\  
\hline
\end{tabular}
\end{center}
\caption{Sensitivity of LC and LHC to the presence of a strong interaction component in W+W- for different luminosities, energies and assuming the presence of a $\rho$ resonance or without it (LET).\label{tab:smtab}} \vspace{0.2cm}
\end{table}

\vspace{3mm}

In case there is no resonance in the J=1 and I=1 channel, one could alternatively use the channels   e$^+$e$^-$   $\rightarrow\nu\bar{\nu}$W$^+$W$^-$ or $\gamma\gamma\rightarrow$W$^+$W$^-$ to access to other quantum numbers.  \par

One can draw similar conclusions in the language of triple gauge couplings, TGC, which should also manifest 
deviations in this scenario. One has 5 TGC preserving parity, custodial symmetry (to avoid effects on PM).
3 are measured at GigaZ and in  e$^+$e$^-$ $\rightarrow$W$^+$W$^-$, 2 with  e$^+$e$^-$ $\rightarrow\nu\bar{\nu}$W$^+$W$^-$.\linebreak
\noindent
These couplings can be expressed as:
$$ \alpha=\frac{\Lambda^2_{EWSB}}{\Lambda^2}  $$
To be significant, a limit on $\Lambda$ should be above $\Lambda_{EWSB}$. This condition is satisfied for the first 3 couplings (also true for the other 2) with LC as shown in figure 12. Note that there is a significant improvement on the determination of these parameters with the input of GigaZ.


\begin{figure}[h]
\begin{center}
\hspace*{-10mm}
\psfig{figure=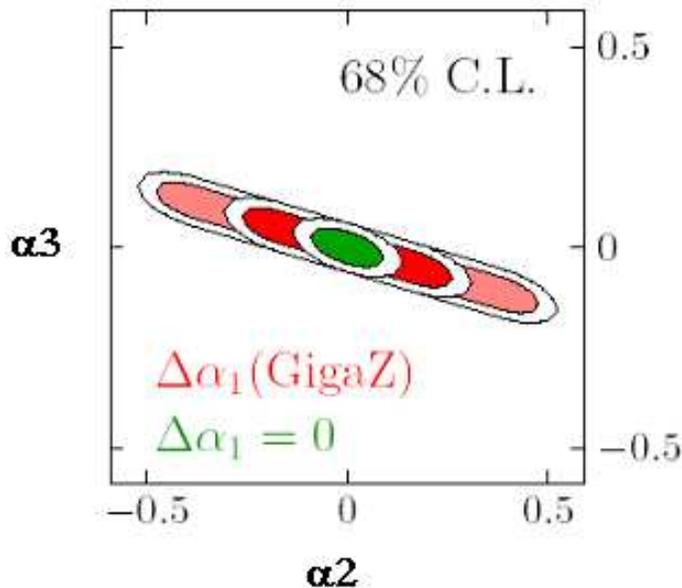,width=10truecm}
\caption{Sensitivity to the triple gauge boson couplings $\alpha_3$ and 
$\alpha_2$ with LC. The central ellipse assumes no error on $\alpha_1$, while the dark red one assumes a precision given by GigaZ.  }
\end{center}
\end{figure}

\section{The SUSY scenario}
 
Although there is very little experimental support for it, SUSY is considered as the leading candidate theory
beyond the SM. On the theoretical side it provides a consistent way to understand EWSB through the Higgs mechanism (no hierarchy problem and EWSB naturally driven by the large Yukawa coupling of the top quark). On the experimental side it allows unification of the 3 forces and it provides natural links to cosmology, in particular a mechanism for generating Dark Matter (DM) in the universe. The g-2 deviation with respect to the SM, which would constitute a precious indication of a SUSY effect, is still uncertain given the 2 contradictory results\cite{davier}obtained
for the hadronic correction. If confirmed if would favor a SUSY solution with light sleptons and light gauginos.\par
\vspace{3mm}
The basic issues for this theory, after it is revealed at LHC or TeVatron, will be: 
\begin{enumerate}
\item To fully confirm the SUSY hypothesis by measuring the spin and couplings (precisely predicted in this theory)of these particles. Recall that, e.g. in the UED framework\cite{UED}, one is able to fake the presence of SUSY but unable to pass above criteria. 
\item To understand the SUSY breaking (SSB) mechanism, for which there is plethora of proposed schemes, none of them 
clearly emerging as preferable to the others\cite{lh}. 
\end{enumerate}
\vspace{0.5mm}  
For pedagogical reasons only, I will use the simplest of SSB schemes, that is mSUGRA in which one can represent the parameter space in two dimensions, in terms of the common scalar and gaugino masses at GUT m0 and M1/2.
tan$\beta$ is a free parameter and $\mu$ is derived in absolute value by imposing EWSB. Figure 13 shows\cite{Ellis} that various experimental conditions, in particular the need to provide the adequate amount of neutral DM and the need to generate
EWSB,
put severe restrictions on these parameters. On can distinguish,in figure 13, four allowed domains:
   \begin{enumerate}
\item The \it{blob} \rm domain, studied so far in LC and LHC studies. This region in which SUSY parameters are moderate is favored by the advocates of \it{Fine Tuning criteria} \rm(FT), meaning that one expects that the SUSY-EWSB generation of the W and Z mass should not result from fine tuned cancellations between large SUSY masses. In this region the DM candidate is a Bino and the presence of light sleptons allows to get sufficient t-channel annihilation 
to keep DM under control.\par 


\begin{figure}[h]
\centering
\psfig{figure=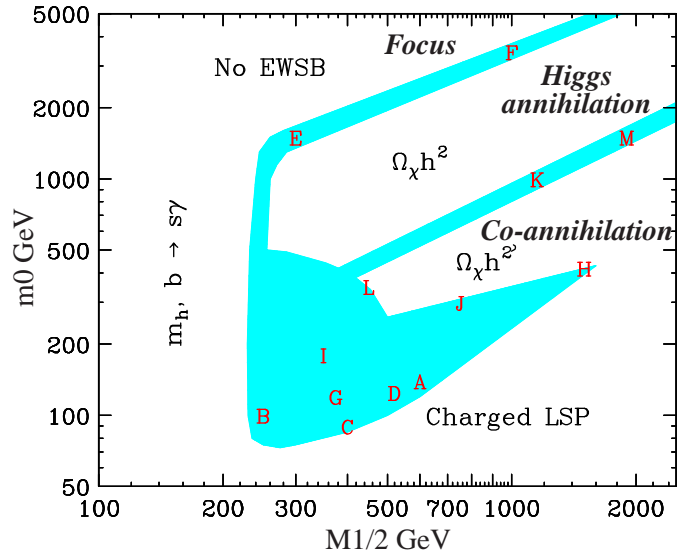,width=9truecm}
\caption{Allowed regions for the m0 (scalar mass) and M1/2 (gaugino mass) parameters within mSUGRA taking into account EWSB, radiative decay of the b quark and the DM constraints. In red are the points which are proposed as benchmarks for LHC and LC. There are 3 regions called 'focus' (large m0 solutions), 'co-annihilation' 
(sleptons almost degenerate with LSP) and 'Higgs annihilation' (LSP about 1/2 the heavy Higgs mass) which allow large values of these parameters.}
\end{figure}

\item In the \it{co-annihilation} \rm domain, the Bino LSP gets heavier and, for DM, one needs to assume that the LSP is almost degenerate in mass with the sleptons, in particular with the stau particle which is usually the 
lightest slepton. The recent results from WMAP, have further restricted this domain as will be discussed later
(see figure 15).
\item The \it{Higgs annihilation} \rm domain corresponds to a region of parameters for which the heavier Higgs particle have a mass close to twice the LSP mass, in such way that efficient s-channel annihilation can occur to reduce the amount of DM. This solution allows to reach high SUSY masses, not only beyond LC reach but even beyond LHC.
\item The \it{focus} \rm domain is remarkable in the sense that it can still claim absence of FT with large values of m0. This is so since the coefficient of m0 in the EWSB equation is minute\cite{feng}. In this scenario, DM can be controlled by noting that the LSP can be a Higgsino if $\mu$ is small, favoring s-channel annihilation. Such a scenario could be quite peculiar since one might only see the 2 first lightest neutralinos and the first chargino, all with masses $\sim\mu$. 
\end{enumerate}

\begin{figure}[h]
\centering
\psfig{figure=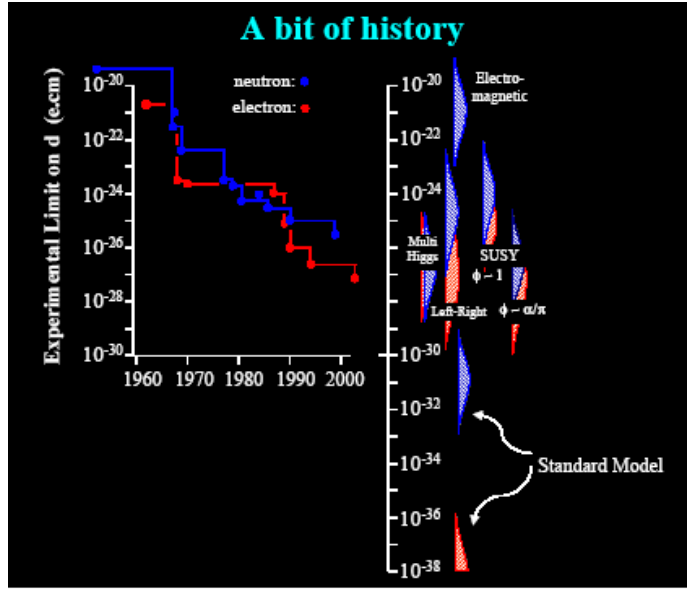,width=9truecm}
\caption{Experimental limit on EDM for neutrons and electrons versus time. On the right hand-side are indicated regions predicted by various theories. The SUSY prediction assumes moderate scalar masses.}
\end{figure}

   Before leaving this part of the discussion, let me point out an important experimental aspect of these arguments on the LSP sector. In scenario 2 and 4 there could be a degeneracy in mass between the sleptons and the LSP and between the gaugino themselves. This conclusion, simply expressed within mSUGRA is certainly quite general\cite{mad} and deserves attention. As shown at LEP2, techniques can be used to detect the charginos in such cases given
 the large production cross-section involved. 
This might not be the case for the slepton with a  much reduced cross-section.

\subsection{The flavor sector}

Previous discussion could lead to the conclusion that nature will choose the less FT solution, that is the 'blob' sector. This conclusion does not take into account the so-called \it{flavor problem} \rm of SUSY. 

This problem extends to the various observables where experimental information puts very severe limitations on SUSY parameters: FCNC constraints, CP violation in the K sector, 
EDM limits for electrons and neutrons (in rapid progress\cite{sauer} as shown in figure 14) and proton lifetime limits. Several authors propose to avoid these constraints by setting the scalar masses, at least for the first two families, to very high value. This could happen in the framework of the 'focus' scenario without drastic FT. \par
There are other ways to avoid some of the problems and restore the 'blob' scenario. One can for example postulate a hidden symmetry (of the Left-Right type as in ref\cite{babu} ) which forces the SUSY phases to 0, or assume a cancellation mechanism\cite{cancel} which reduces the importance of CP violation in the EDM process. The later seems to also imply some kind of fine-tuning given that one already needs to assume a very small phase for the $\mu$ term. \par
\subsection{The 3 scenarios}    

From previous discussion and for the sake of simplicity, one may foresee 3 types of scenarios representing different challenges for a LC:
\begin{enumerate}
\item All SUSY particles are very heavy except the lightest Higgs boson h and, possibly, the two lightest neutralinos and 
the lightest chargino accessible at LC and the gluino at LHC. This would correspond to a 'focus' type solution. 
The LSP could be a Higgsino with $|\mu|<$M$_1$ or a Wino if the usual inequality M$_1<$M$_2$ is violated. In both cases there would be a substantial s-channel annihilation. The three lightest gauginos are almost degenerate 
in mass but, as shown at LEP2, observable at LC. 
\item Same as 1 but, in addition, the third generation of scalar particles is light. Then co-annihilation can take place between the LSP and the stau particle. As pointed in\cite{olive} , the new results from WMAP would require a very close degeneracy and therefore the stau would be extremely hard to observe. The good news however is that the WMAP results imply that the LSP cannot be heavier than $\sim$ 500 GeV and therefore would fall within the range of observation of a LC. 
\item The 'blob' scenario which allows a wide range of possibilities subject however to the Higgs mass limit of LEP2 
and, possibly, to the g-2 measurement from BNL. Figure 15 indicates a recent update\cite{olive} 
illustrating these features. 
\end{enumerate}
    
\begin{figure}[h]
\centering
\vspace{-5mm}
\psfig{figure=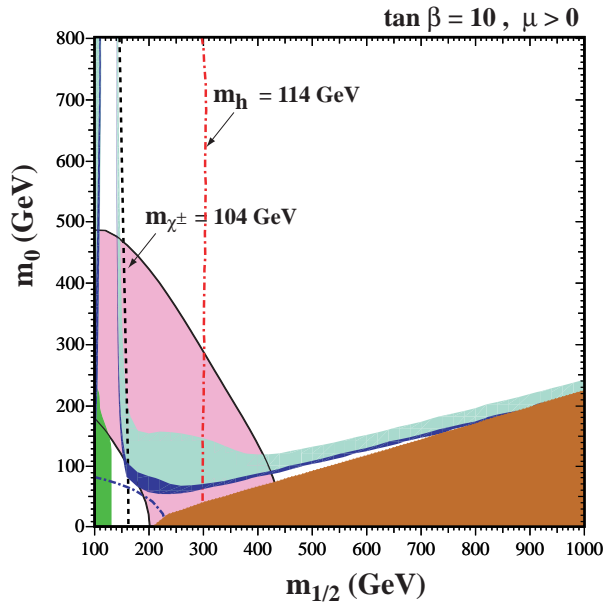,width=8truecm}
\caption{Allowed regions within mSUGRA assuming $\mu>$0 and tan$\beta$=10. Indicated also is the mass limit from LEP2 on charginos and on the Higgs boson mass. The small bands correspond to DM solutions with (dark blue) and without (pale green) WMAP derived results. The pink band is favoured by an interpretation of the  g-2 
results, still under investigation.}
\end{figure} 
\newpage

\subsection{DM at LC}
If within reach, LC will accurately measure the mass of the LSP and its couplings, in other words its Higgsino/Wino/Bino content.\par
\noindent
There are various tools to perform such an analysis, threshold scans, polarization asymmetries which I will not describe here but can be found in\cite{tdrt}. \par
These measurements will constitute an input for cosmology as will be illustrated shortly. They will also allow
to precisely interpret the non-accelerator searches on primordial DM performed with various techniques\cite{jesus}. \par
Taking point B within mSUGRA (see figure 13), a case treated with LHC inputs in reference\cite{ellisup}, I have estimated the accuracy on
$\Omega_{DM}$h$^2$ from SUSY using the code MicroMEGAs\cite{micro}, given that a LC can measure the LSP and 
$\tau$ slepton masses with accuracies 0.1 GeV and 0.6 GeV respectively\cite{tdrt}. Any significant discrepancy with cosmology may reveal extra sources of DM (e.g. axions or very heavy objects produced in the early phases of the universe). 
\par
It is also worth recalling\cite{tdrt} that the chargino and neutralino measurements, if there is a Bino component, provide an indirect sensitivity on selectrons and sneutrinos masses which are precise up to masses $\sim$1 TeV, that is well above the reach of a TeV collider. 

\vspace{3mm}
\begin{table}[h]
\begin{center}

\vspace{0.2cm}
\begin{tabular}{|c|c|}
\hline WMAP
&
 7\%  \\
\hline
LHC & $\sim$ 15\% \\
\hline Planck &$\sim$ 2\%\\
\hline LC &$\sim$ 3\% \\   
\hline
\end{tabular}
\end{center}
\caption{Relative precision on DM obtained by cosmology in the present phase (WMAP) and foreseen (Planck), to be compared with the accuracy foreseen on its determination at LHC and at LC assuming point B as defined in figure 13.\label{tab:smtab}} 
\end{table}

\vspace{3mm}
\section{ LC and the GUT scale}

As already stated, understanding the origin of SSB will be the major goal of our field once SUSY is discovered at TeVatron or LHC. To reach this goal, it will be essential to measure SUSY masses and couplings precisely and in a 
model independent framework. As an example, in the gaugino sector, LC should allow to measure precisely M1 and M2 while M3 is provided by LHC from the gluino mass. These three quantities would be extrapolated to the GUT scale and provide a crucial test of SSB. Figure 16, taken from\cite{zerw}, gives an example of such a confrontation in the case of 
a string inspired model. \par
\vspace{8mm}
\begin{figure}[h]
\centering
\psfig{figure=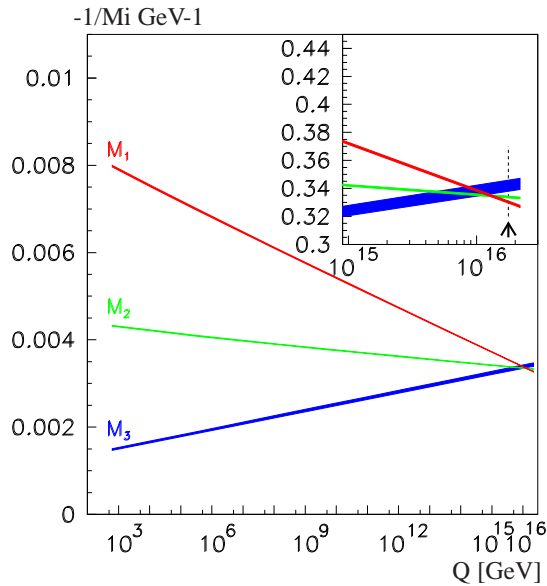,width=8truecm}
\vspace{2mm}
\caption{Evolution of the gaugino masses up to the GUT scale with a window showing the breaking of universality due
to loop corrections in a string inspired model.}
\end{figure} 

 One can observe that while the convergence is satisfied, it occurs at a mass below the GUT scale given by the couplings. This shift is due to loop corrections present in such a scheme and 
the LC accuracy is needed to observe this subtle effect present in string theories\cite{binet}. 
This example shows how a LC may turn out to be crucial to understand the underlying fundamental theory. \par

\newpage
Moreover one can observe in figure 16 the unmatched accuracy of M3. Recently\cite{weig}it has been shown that this accuracy can be greatly improved if LHC can use the LC information, in particular the mass determination of the LSP. One could therefore reduce the error given in figure 16 by a factor 3.     \par    
This example illustrates the potential of a LC+LHC combination to reach a very ambitious goal, pioneered by LEP/SLD: 
explore the GUT/string scale to reconstruct the fundamental parameters of an underlying theory which no accelerator will 
ever be able to test directly.  
Studies of this type are systematically being done in the framework of LHC/LC collaborative studies\cite{weig}. 
\section{ Summary: Why do we need a LC ?}
From this very incomplete description of the prominent topics which will be addressed by a LC, one can derive the following list of goals:
\begin{enumerate}
\item To provide the full picture on an SM/MSSM Higgs.
\item To provide an answer on the issue on EWSB even in more difficult or unexpected situations, e.g. a reduced 
cross-section, a heavy Higgs or no Higgs 
scenario.
\item To access to the SSB mechanism combining LC and LHC measurements and, even more ambitiously, to access to GUT scale and to the parameters of the
underlying string theory. 
\item To predict very precisely, within SUSY, the amount of DM and confront it to the observation by cosmology
\item  To interpret unambiguously an unexpected discovery at LHC, e.g. a Z' or a Kaluza Klein excitation.
\item To estimate mass scales beyond LC/LHC reach:\par
- Observing deviations on PM translated, e.g., into
a heavy Higgs, a heavy sfermion or a Z' mass  \par
- Testing the theory at the quantum level and eventually predicting new mass scales as has already occurred with 
LEP/SLD/TeVatron for the Higgs mass \par
These various examples indicate how PM at a LC may reveal new frontiers in energy which would call for an upgraded LHC or for a new generation of colliders like CLIC or VLHC.    
\end{enumerate}

\section*{Acknowledgements}
It is a pleasure to thank and congratulate the team of this Conference 
for the excellent organisation of the meeting.

\clearpage
\section*{DISCUSSION}

\begin{description}
\item[Chang Kee Jung] (Stony Brook): 
What is the most current MGUT apparent value with an error?
What is the MGUT value newly derived by this particular string model shown in your talk?
\item[Francois Richard:] In the plot shown, the gaugino masses unify at about 10$^{16}$ GeV in contrast with the coupling constants which unify at twice this value. The claim is that such effect can be seen at a LC. In this particular case the effect originates from loop corrections within a string inspired theory.
\item[Thomas Germann] (Zurich University):
You mention the option of using transverse polarization. What are the physics observables to be probed with this? \par
\item[Francois Richard:]
I had in mind the possibility of measuring polarization asymmetries which are insensitive to vector contributions but can reveal tensor terms originating from low scale gravity contributions. This topic has been recently advocated by T. Rizzo et al., and they find a sensitivity of this method up to very high mass scales. 
\item [Andreas Kronfeld] (FNAL):
If I understood your remarks on dark matter correctly, cosmology could
determine the total component of dark matter at the 2\% level. Meanwhile,
LC could determine the contribution of a single identified component
(e.g. the LSP) with similar precision. Then you could imagine
DM$_{Total}$= (30.0+/-0.5)\% and DM$_{LSP}$= (20.0+/-0.5)\%. That would be
fascinating.  Did I understand this correctly?
\item[Francois Richard:] Yes, there is no reason that the two numbers coincide 
but it matters that the precisions are similar for an optimal comparison.  
\end{description}

\end{document}